\documentclass[pre,twocolumn,superscriptaddress,showpacs,amsmath,amssymb,floatfix,amstex]{revtex4-1}

\usepackage{units}
\usepackage{psfrag}

\usepackage[dvips]{graphicx}
\usepackage{color}

\newcommand{\mean}[1]{\langle #1 \rangle}

\begin{document}

\title{Finite-size-induced transitions to synchrony in oscillator ensembles with nonlinear
global coupling}

\author{Maxim Komarov} 
\affiliation{Department of Physics and Astronomy, University of Potsdam,
  Karl-Liebknecht-Str 24, D-14476, Potsdam, Germany}
  \affiliation{Department of Cell Biology and Neuroscience, University of California Riverside,
  900 University Ave. Riverside, CA 92521, USA}
\affiliation{Department of Control Theory, Nizhni Novgorod State University,
  Gagarin Av. 23, 606950, Nizhni Novgorod, Russia}
\author{Arkady Pikovsky} 
\affiliation{Department of Physics and Astronomy, University of Potsdam, 
  Karl-Liebknecht-Str 24, D-14476, Potsdam, Germany}
  \affiliation{Department of Control Theory, Nizhni Novgorod State University,
  Gagarin Av. 23, 606950, Nizhni Novgorod, Russia}
\date{\today}

\begin{abstract}
We report on finite-sized-induced transitions to synchrony 
in a population of phase oscillators coupled via a nonlinear mean field,
which microscopically is equivalent to a hypernetwork organization of interactions. 
Using a self-consistent approach and direct numerical simulations, we argue that a
 transition to synchrony occurs only for 
finite-size ensembles, and disappears in the thermodynamic limit.
For all considered setups, that include purely 
deterministic oscillators with or without heterogeneity 
in natural oscillatory frequencies, and an ensemble of noise-driven identical 
oscillators, we establish scaling relations describing the order parameter
as a function of the coupling constant and the system size.
\end{abstract}

\pacs{05.45.Xt, 05.45.-a}

\maketitle
Collective synchronization phenomena are abundant in complex 
nonlinear systems, and onset of synchrony can be 
typically treated as a nonequilibrium phase 
transition. The Kuramoto model~\cite{Kuramoto-75,*Acebron-etal-05} 
of globally coupled phase oscillators is the simplest paradigmatic system, 
where
this transition can be explored nearly in full 
details~\cite{Strogatz-00,*Ott-Antonsen-08}, also a relation to equilibrium 
transitions is well studied~\cite{Gupta-Campa-Ruffo-14}. This model
is universally applicable for ensembles of weakly coupled oscillators,
possessing harmonic phase sensitivity (like, e.g., Josephson 
junctions~\cite{Wiesenfeld-Colet-Strogatz-96}).  
In many cases one, however, needs to
go beyond such a simple setup, allowing for couplings 
that include higher harmonics, this is relevant for electrochemical
oscillators~\cite{Kiss-Zhai-Hudson-05} and $\varphi$-Josephson 
junctions~\cite{Goldobin_etal-11}. Moreover, as 
suggested in~\cite{Rosenblum-Pikovsky-07} and experimentally realized 
in~\cite{Temirbayev_et_al-12,*Temirbayev_et_al-13}, coupling terms can be
nonlinear functions of the order parameters (mean fields).

In this letter we describe nontrivial properties of the synchronization transition
in a model with simple nonlinear coupling, where the coupling is at the
second harmonics of the phase, but is proportional to the 
the square of the Kuramoto order parameter. 
We will show that such 
an interaction  on a microscopic level represents a fully connected hypernetwork.
By performing the analysis in the thermodynamic limit,
we will demonstrate, that for deterministic ensembles the asynchronous state with a 
uniform distribution of the phases never loses stability; for noisy oscillators it is possible
to show that such an asynchropnous state is the only stationary solution. Nevertheless, direct 
numerical simulations of finite pupulations
yield partially synchronous regimes. These regimes can be 
called finite-size-induced one; the main goal of this paper is to clarify their
nature. Our main tool is the analysis of scaling of partial synchrony
with the system size. We will establish scaling properties of this 
finite-size-induced transition for different setups: identical purely deterministic
oscillators, identical noisy oscillators, and deterministic non-identical ensemble (the 
latter setup is mostly close to the Kuramoto model). From these scaling properties
it follows, that in the thermodinamic limit the characteristic value of the observed order parameter 
tends to zero, while the critical value of the coupling strength at which partial synchrony 
is observed, diverges. Thus, this transition to synchrony can be called finite-size-induced one.


Let us start with formulation of general phase equations 
for an  ensemble of nonlinear oscillators coupled via mean fields.
In the limit of weak coupling (or weak external forcing), 
the dynamics of each oscillator 
can be described by its phase $\phi(t)$ via the phase response function $S(\phi)$:
\[
\dot\phi=\Omega+\omega+S(\phi)F,
\]
here $\Omega$ stands for the population mean natural frequency and $\omega$ 
is an individual deviation from the mean.
The term $F$ is an overall force produced by mean field coupling,
i.e. it is a general function of mean fields (Daido order 
parameters~\cite{Daido-93,*Daido-96})
 $Z_k=\mean{e^{ik\phi}}$
(here $\mean{}$ means averaging over the ensemble) which can be represented as an
expansion $F=\sum_{k,m} h_{k,m} (Z_k)^{m} $ (where $h_{k,m}$ 
are constants).
Next, let us introduce a slowly varying phase $\varphi(t) = \phi(t)-\Omega t$ 
and the corresponding slow order 
parameters $\bar{Z}_k=\mean{e^{ik\varphi}} = Z_k e^{-i k \Omega t}$. 
Using a Fourier representation of the phase response function  
$S(\phi) = \sum_n s_n e^{-in\phi}$ we get the following general
equation for the slow phase:
 \begin{equation}
 \label{eq:ph1}
\dot\varphi=
\omega+\sum_{n,k,m} s_ne^{-in\varphi}h_{k,m}\bar{Z}_k^me^{i(km-n)\Omega t}.
\end{equation}
Performing an averaging of the equation (\ref{eq:ph1}) 
over the fast time scale $\Omega^{-1}$ is equivalent to keeping
only terms on the r.h.s. that do not contain explicit time dependence 
$\sim e^{ij\Omega t}$, i.e. we have to set $n=km$:
 \begin{equation}
 \label{eq:ph2}
\dot\varphi= \omega+\sum_{k,m}s_{km}e^{-ikm\varphi}h_{k,m}\bar{Z}_k^m.
\end{equation}
This is the general form of the phase equation derived 
for a weakly coupled oscillatory ensemble with generically nonlinear
(due to terms with $m>1$)
mean-field interaction 
(cf.~\cite{Rosenblum-Pikovsky-07}).
Recall that the simplest case where $n=k=m=\pm 1$ gives us the 
term $\sim \bar{Z_1}e^{-i\varphi}$, and \eqref{eq:ph2} reduces to 
the standard Kuramoto-Sakaguchi model~\cite{Sakaguchi-Kuramoto-86}.
Considering further terms with $n=km=\pm 2$, we get a 
generalized bi-harmonic coupling 
$\sim s_1 h_{1,1} \bar{Z}_1e^{-i\varphi}+(s_2 h_{2,1} \bar{Z}_2+
s_2 h_{1,2} \bar{Z}_1^2)e^{-i2\varphi}$ on the r.h.s of the system (\ref{eq:ph2}).
The case where the bi-harmonic coupling depends linearly on the order 
parameters ($h_{1,2}=0$) was extensively studied in 
Refs.~\cite{Komarov-Pikovsky-13a,*Komarov-Pikovsky-14,*Vlasov-Komarov-Pikovsky-15}.

In this letter we focus on the effects produced by purely
nonlinear second-harmonic coupling $s_1=h_{2,1}=0$ (see 
\cite{Skardal-Ott-Restrepo-11} for the analysis
of linear second-harmonic coupling $s_1=h_{1,2}=0$) 
and show that it is responsible for 
finite-size induced transitions to synchrony with nontrivial scaling
on the ensemble size. We will demonstrate that while synchrony disappears in 
the thermodynamic limit, it is observed for finite ensembles.
Thus, throughout this paper we consider the following model, where also external
noise is taken into account for completeness:
 \begin{equation}
 \label{eq:main}
\dot\varphi_k= \omega_k+\varepsilon R^2\sin(2\Theta-2\varphi_k)+\sqrt{D}\eta_k(t),\ \ \ k=1,...,N.
\end{equation}
Here we denote $Re^{i\Theta}=\bar{Z}_1=N^{-1}\sum_{k=1}^{N}e^{i\varphi_k}$, 
$N$ is the size of population, and the noise is Gaussian delta-correlated 
$\mean{\eta_k(t)\eta_m(t-t_0)}=\delta_{km}\delta(t-t_0)$. Qualitatively,
nontrivial features of this model can be understood as follows. Because the interaction is proportional
to the second harmonics of the phase $\sim \sin 2\varphi$, it supports formation of two clusters, with phase
difference $\pi$. However, the coupling term is determined by the order parameter $R$ which
vanishes if two symmetric clusters are formed and is non-zero only due to asymmetry of the clusters.
This asymmetry, as we shall demonstrate below, is due to finite-size fluctuations at the 
initial stage when the clusters are formed from the disorder.

Before proceed with the main analysis, we discuss physical 
relevance of the model (\ref{eq:main}).
First, the purely second-harmonic coupling ($s_1=0$)
appears when the force acts on nearly harmonic oscillators
parametrically (a typical example here are 
mechanical pendula suspended on a vertically oscillating
common beam). Another situation where the second harmonic
in $S(\phi)$ dominates is that of period-doubled oscillations. 
Noteworthy, due to nonlinear coupling model (\ref{eq:main}) represents
a hypernetwork~\cite{Wang_etal-10,*Bilal-Ramaswamy-14} of oscillators. 
Indeed, substituting the expression
for the mean field in (\ref{eq:main}), one can see that the 
microscopic coupling
terms can be written as $\sim \sin(\varphi_l+\varphi_m-2\varphi_k)$.
This means that effective 
interactions are not pairwise (as in the standard Kuramoto model
and its numerous generalizations)
but via triplets; this is exactly the definition of a 
hypernetwork coupling structure.  

Furthermore, it is worth discussing the r\^ole of different order parameters in the problem.
The order parameter $\bar{Z}_1$ governs the force acting on the oscillators and is therefore of major
importance. Because this force contains the second harmonics only ($\sim\sin 2\varphi_k$),
the appearing order is of ``nematic'' type and corresponds to large absolute values of the order parameter
$\bar{Z}_2$; at these states the order parameter $R=|\bar{Z}_1|$ may be small.  
In the disordered, asynchronous states both order parameters $\bar{Z}_{1,2}$ 
are small.

Linear stability analysis of a disordered state ($R=0$) in model~\eqref{eq:main}
is straightforward, beacuse coupling is nonlinear in $R$ 
and thus does not contribute.
The solution is the same as for the Kuramoto model with vanishing 
coupling~\cite{Strogatz-Mirollo-91,*Crawford-Davies-99}:
the disordered state is either neutrally (without noise $D=0$), or asymptotically
stable. Thus, all the transitions described below are due to
nonlinear and finite-size 
effects.

We start with the simplest case where all oscillators have 
identical frequencies ($\omega_k=0$) and are not affected by 
noise ($D=0$).
In this case, for any $Re^{i\Theta}=const\neq 0$ there are two stable positions
for the phases: $\varphi_1=\Theta$ and $\varphi_2=\Theta+\pi$.
Any distribution $(n_1,n_2)$ with $n_1>N/2$ oscillators in the first state is possible,
with order
parameter 
\begin{equation}
R=2n_1/N-1\;.
\label{eq:rvsn}
\end{equation}
Only the symmetric distribution
with $n_1=N/2$  is not a solution, because the mean field vanishes.
Therefore, in the thermodynamic limit $N\to\infty$, the stationary 
two-cluster distributions
can be written as
\begin{equation}
\rho(\varphi) = S\delta(\varphi-\Theta)+(1-S)\delta(\varphi-\Theta+\pi),
\label{eq:indic}
\end{equation}
with an arbitrary indicator constant $S\in \big(\frac{1}{2},1\big]$, the
order parameter is $R=2S-1$ and $\Theta$ is arbitrary.

\begin{figure}
\centering
\includegraphics[width=\columnwidth]{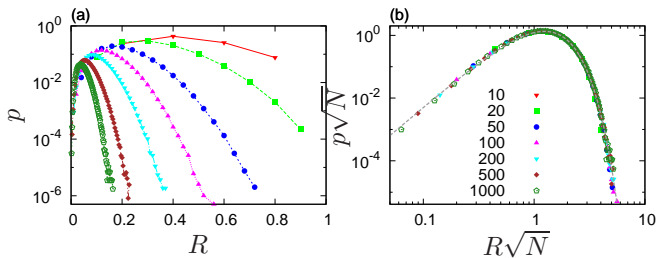}
\caption{(color online) The histograms of the two-cluster states which result from the disordered initial 
conditions $R\approx 0$, for different
ensemble sizes from $N=10$ to $N=1000$. 
Panel (b) shows scaling $p\sim N^{-1/2}F(R N^{1/2})$;
dashed line is a fitting curve $F(x)=5.2 x^3 \exp[-1.45 x^{3/2}]$.
Theoretical derivation of this scaling relation remains an unsolved problem.
}
\label{fig:fig1}
\end{figure}

A nontrivial question here is: which of possible synchronous states 
establishes if one starts from a fully disordered initial configuration
with uniformly distributed initial phases $0\leq \varphi_k<2\pi$.
Numerically obtained distributions of the final states, for different
sizes $N$, are shown in Fig~\ref{fig:fig1}(a).
Here the order parameter $R$ 
can attain only discrete values according to \eqref{eq:rvsn}, and $p$ are probabilities of these states.
Remarkably, these distributions collapse
perfectly after rescaling $R\to R\sqrt{N}$,
$p\to p\sqrt{N}$, as  is shown in Fig.~\ref{fig:fig1}(b). This means that 
the stationary
order parameter scales as $R\sim N^{-1/2}$, i.e. it disappears in the
thermodynamic limit. To this scaling corresponds also the scaling of the
characteristic transient time from initial 
disorder to a final synchronous configuration:
as one can see from \eqref{eq:main}, this time is $\sim R^{-2}\sim N$, what is 
confirmed by numerics (not shown).  

Next, we consider the case when the oscillators have 
different natural frequencies $\omega_k$ and are not affected by noise $D=0$.
We assume a Gaussian distribution 
$g(\omega)= (2\pi)^{-1/2} e^{-\omega^2/2}$ 
(without loss of generality the width of the distribution is set to 1 and 
the mean frequency to 0).
First, following 
Refs.~\cite{Komarov-Pikovsky-13a,*Komarov-Pikovsky-14,*Vlasov-Komarov-Pikovsky-15}, 
we find stationary solutions 
in the thermodynamical limit
by virtue of a self-consistent scheme. We will see, that although the analysis
of the thermodynamic limit does not provide a transition to partial synchrony, it allows us
to find states close to that observed in simulations of finite systems.
For the sake of brevity of presentation
we restrict to symmetric, non-rotating solutions only, then without loss of generality
one can set $\Theta=0$. For such states
the conditional distribution density $\rho(\varphi |\omega)$ is stationary,
and the order 
parameter can be defined as follows:
\begin{equation}
\label{eq:self_cons_order_p}
R = \iint d\omega\, d\varphi\, g(\omega)\rho(\varphi |\omega) \cos\varphi.
\end{equation}

\begin{figure}
\centerline{\includegraphics[width=0.49\columnwidth]{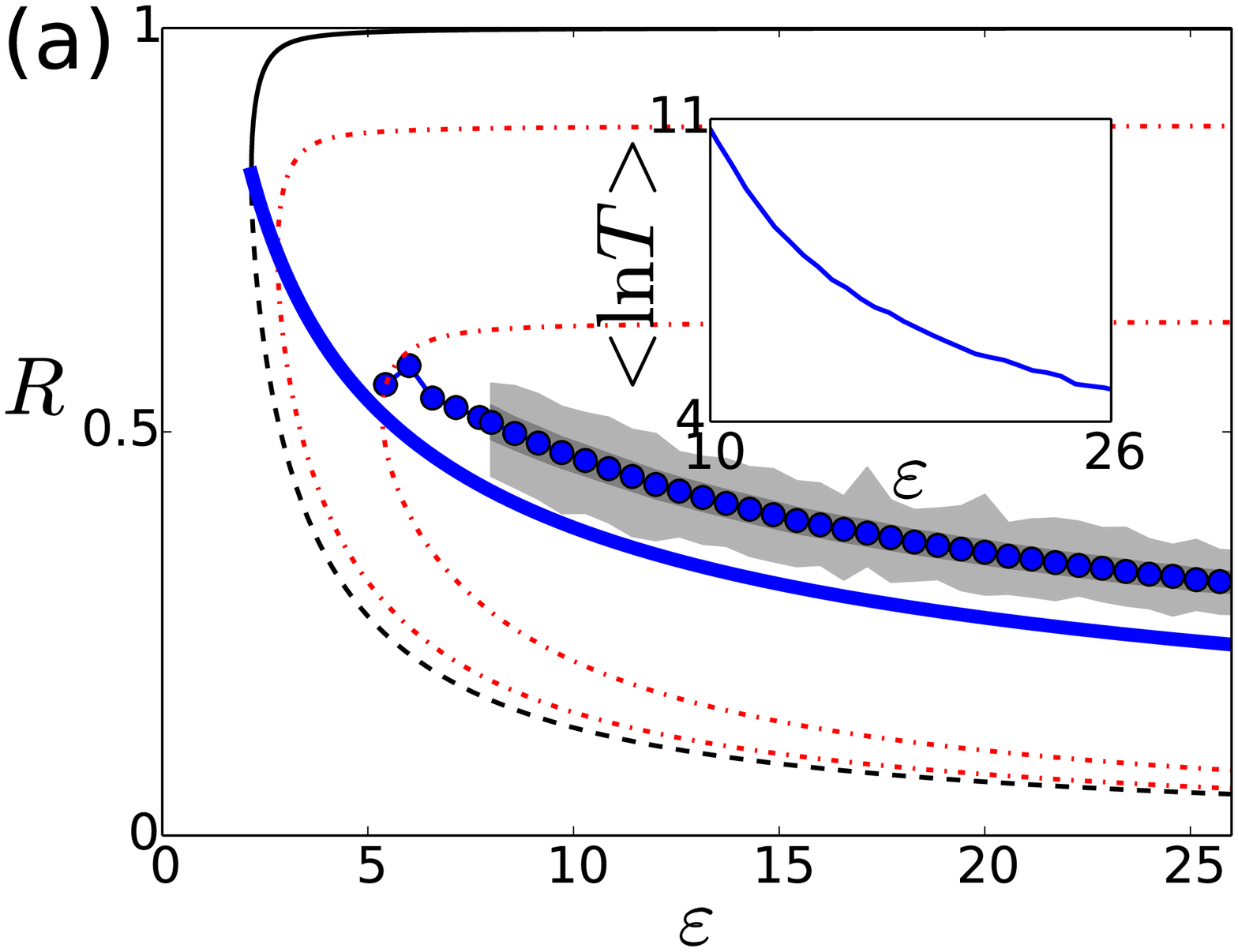} 
\includegraphics[width=0.49\columnwidth]{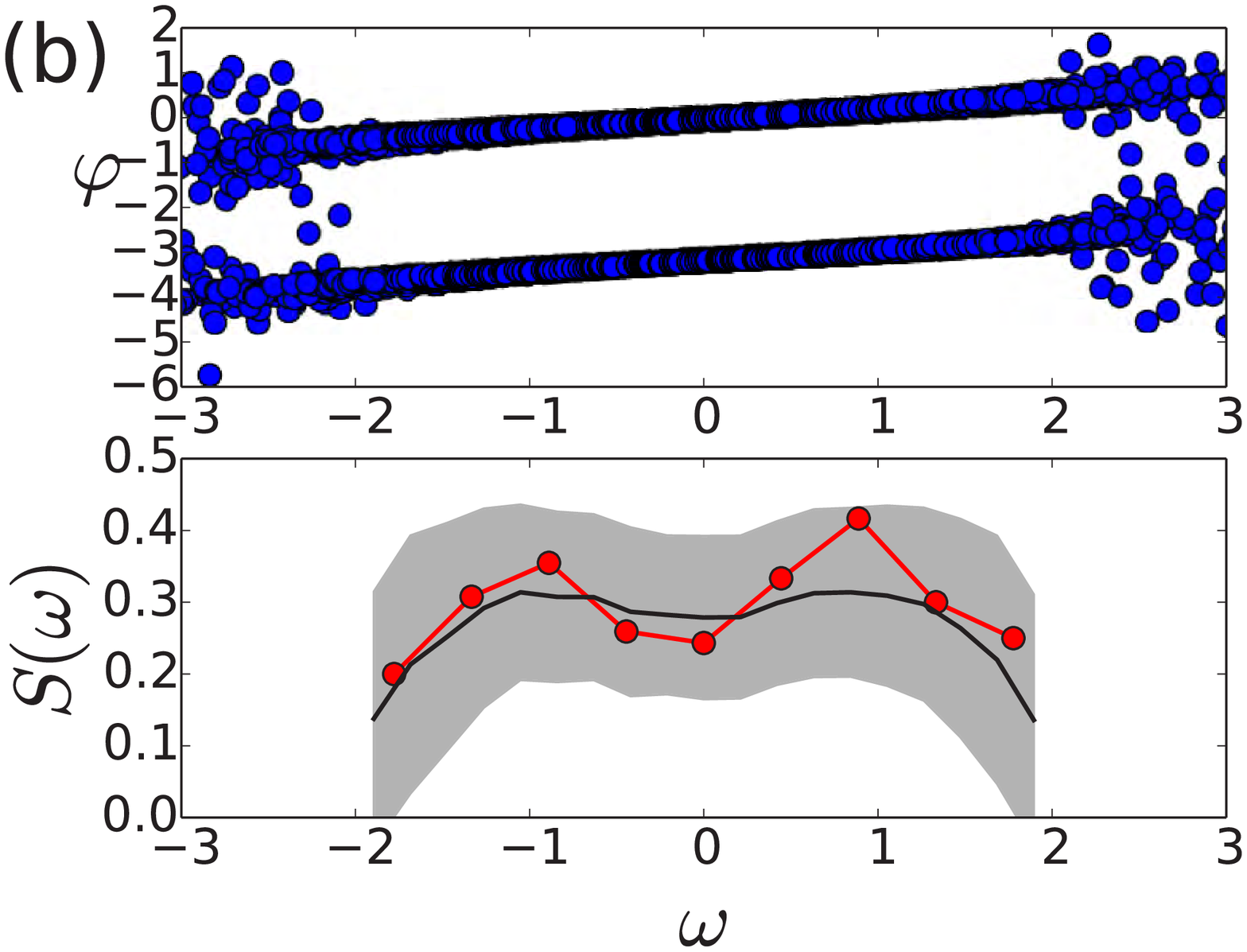}}
\centerline{\includegraphics[width=0.49\columnwidth]{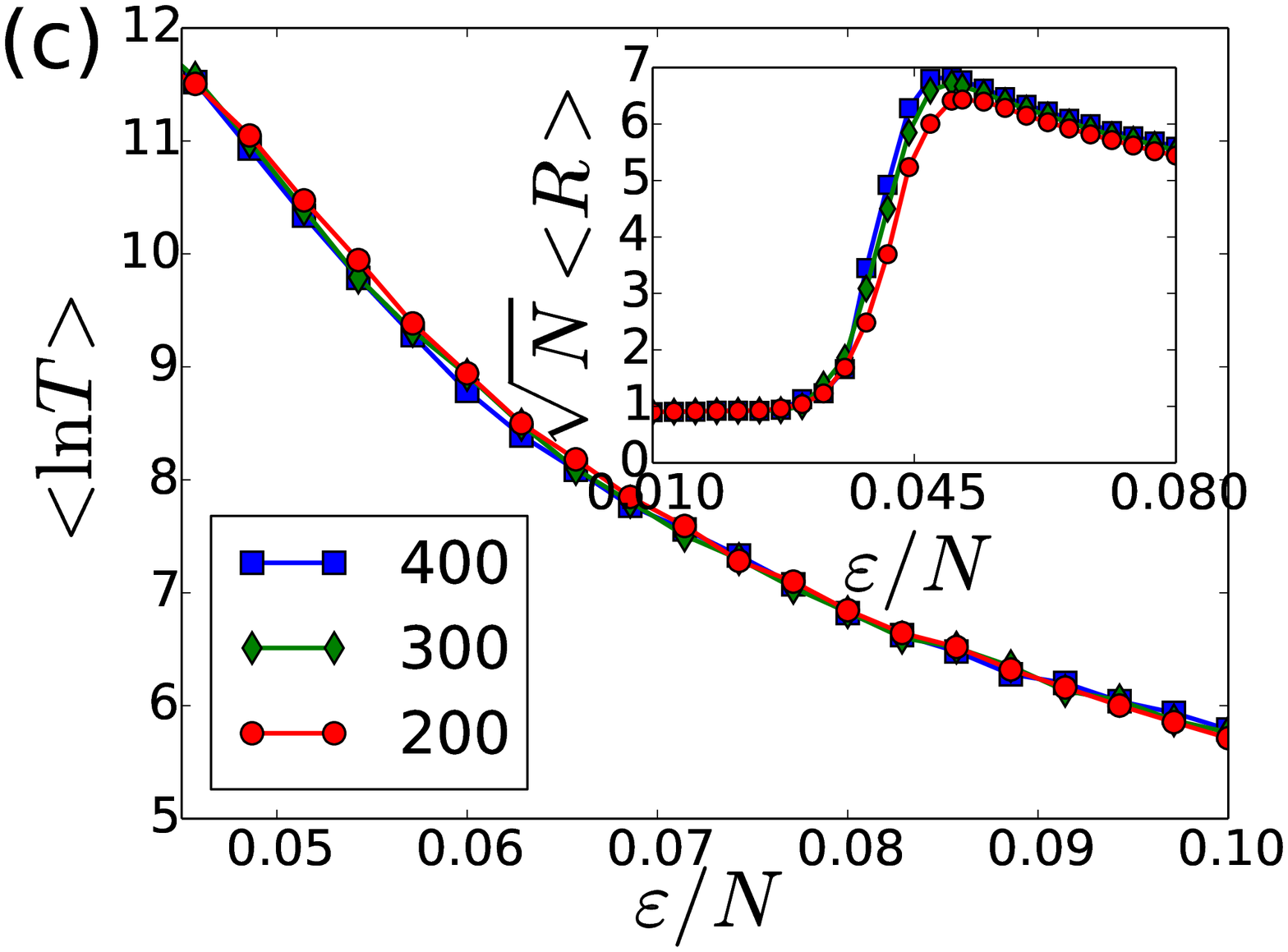} }
\caption{(color online) (a) Black solid and dashed curves: 
solution of the self-consistent equations (\ref{eq:self_cons}) for $\sigma=1$. 
Blue bold curve - the bifurcation 
line where two branches of synchronous solutions appear for different $\sigma$. 
Blue markers: mean value of $R$ obtained from direct numerical simulation of 
(\ref{eq:main}) 
with $N=200$ and initial conditions in the incoherent state.
Dotted red lines denote solution of self-consistent equation (\ref{eq:self_cons}) 
for $\sigma=0.94$ and $\sigma=0.82$.
The inset show the averaged lifetime $T$ of the incoherent state for the 
finite-size ensemble.
(b) The top panel shows coordinates of phase $\varphi$ as a function of internal frequency $\omega$ in the stationary synchronous state (the data is shown for 1000 different simulations at $\varepsilon=15$, $N=200$). The bottom panel depicts function $S(\omega)$ in the range where the oscillators are locked and form two distinct stable branches. The black solid curve stands for averaged values of $S(\omega)$, the gray area denotes the standard deviation, and the red markers depict one particular realization. In order to calculate $S(\omega)$, the range $\omega\in [-2,2]$ was split into $m$ subintervals. In each subinterval we calculate the ratio of oscillators, located at different branches. For averaged black curve we use $m=20$, for red curve $m=10$.
(c) The dependence of the averaged lifetime $T$ of the incoherent state 
is shown for different system sizes $N$. The inset depicts the averaged 
value of order parameter $R$ obtained from finite-size finite-time 
($T_{max}=10^6$) simulations of the system (\ref{eq:main}).
}
\label{fig:fig2}
\end{figure}

It is convenient to introduce an auxiliary 
parameter $A=\varepsilon R^2$ (the overall amplitude of the 
coupling function) and the rescaled 
frequency $x=\omega/A$. 
It easy to see from (\ref{eq:main}) that all the oscillators can be divided 
into those locked by the mean field ($|x|<1$) and the unlocked 
(rotating) ones $(|x|>1)$. The distribution of the latter ones 
is inversely 
proportional to the phase 
velocity $\rho_u(\varphi |x)=C|A(x-\sin(2\varphi))|^{-1}$ 
(here $C$ is a normalization constant) and, because of the symmetry, 
it does not contribute to the order parameter 
in (\ref{eq:self_cons_order_p}).
The distribution of the locked oscillators $\rho_L(\varphi |x)$ is in 
fact the same as in \eqref{eq:indic}, but frequency-dependent:
\begin{equation}
\label{eq:self_cons_distr_fun}
\begin{aligned}
\rho_L(\varphi |x) &= S(x)\delta(\varphi - \Phi)+(1-S(x))\delta(\varphi-\Phi-\pi)
\end{aligned}
\end{equation}
where $\Phi(x)={\arcsin(x)}/{2}$. 
Similar to the case of identical oscillators, the 
indicator function $S(x)$ is arbitrary (due to assumed
symmetry we restrict ourselves to the case $S(x)=S(-x)$,
asymmetric functions lead generally to rotating solutions), it 
describes redistribution of oscillators between two 
stable branches $\varphi=\Phi(x)$ and $\varphi=\Phi(x)+\pi$.
Below we consider the simplest case of constant indicator function 
$S(x)=\sigma\in \big(\frac{1}{2},1]$.
In order to get the final closed self-consistent scheme, 
we need to substitute the distribution function (\ref{eq:self_cons_distr_fun}) 
into the definition of the order parameter (\ref{eq:self_cons_order_p}).
This yields the 
order parameter $R$ as a function of the coupling constant 
$\varepsilon$ in a parametric (in dependence on auxillary parameter $A$) 
analytic form:
\begin{equation}
\label{eq:self_cons}
\begin{aligned}
R &= A(2\sigma-1) H(A),\ \ \ \varepsilon = A R^{-2},\\
H(A) &= \int_{-\frac{\pi}{4}}^{\frac{\pi}{4}}2\cos\varphi\cos2\varphi 
g(A\sin2\varphi)d\varphi.
\end{aligned}
\end{equation}
Figure~\ref{fig:fig2}(a) illustrates stationary solutions 
$R(\varepsilon)$ at different indicator constants $\sigma$.
The black curves denote the main solution at $\sigma=1$.
At a critical coupling $\varepsilon_c\approx 2.17$
two branches, stable (black solid) and unstable (dashed line)
arise (stability is determined by direct finite-size numerical simulations).
Note that these lines are separated from the disordered solution
$R=0$, although $R\sim \varepsilon^{-1}$ as $\varepsilon\to \infty$
at the unstable branch. Solutions for $\sigma<1$ can be easily 
found from the  rescaling of the main dependence (at $\sigma=1$)
according to~\eqref{eq:self_cons}. So, the curves $R(\varepsilon)$ at $\sigma<1$ 
have qualitatively 
similar structure, however, they are shifted to larger values of 
$\varepsilon$ for smaller values of $\sigma$ (see dotted red lines in Fig.~\ref{fig:fig2}(a)). 
In particular, the critical
points scale as $R_c(\sigma)=R_c(0)\left(\frac{\varepsilon_c(1)}
{\varepsilon_c(\sigma)}\right)^{1/2}$.
This blue bold line in Fig.~\ref{fig:fig2}(a) together with the black solid line
at $\sigma=1$ define the region of possible synchronous solutions,
 characterized by different indicator 
constants $\sigma$.

It is worth mentioning that the incoherent solution $R=0$ exists at any value of 
coupling $\varepsilon$ and it is always stable in the thermodynamical limit.
However, in the finite-size simulations of the system (\ref{eq:main}), we found 
that the incoherent state has a finite lifetime: after a long transient a synchronous
state from the described above solution family (i.e.
between the blue bold and black solid curves in Fig.~\ref{fig:fig2}(a)) sets on, 
and this state remains further stable.
Blue markers in the Fig.~\ref{fig:fig2}(a) denote averaged value of $R$ 
obtained from direct numerical simulations of (\ref{eq:main}) with $N=200$.
The averaging was performed over $\sim 1000$ distinct simulation runs (until
final time $T_{max}=10^6$) with 
disordered initial conditions.
The final state to which the system jumps from the incoherence is 
not always the same and has a deviation range depicted by the 
gray area in Fig.~\ref{fig:fig2}(a). 

A more detailed description of the final synchronous state is presented in the Fig.~\ref{fig:fig2}(b). 
Here the top panel shows dependence of the phases $\varphi$ on the internal frequency $\omega$. 
The area in the center clearly shows two stable branches of locked oscillators.
Outside this area one can see the clouds of points, which depict unlocked oscillators. The unlocked phases rotate in relation to the mean field phase, and, therefore, constitute an asynchronous part of the ensemble.
The bottom panel shows statistics of the function $S(\omega)$, which is calculated in the range $\omega\in [-2,2]$ where oscillators are typically locked to the mean field.
The function has certain profile depicted in the bottom panel of the Fig.~\ref{fig:fig2}(b). 
As one can see, the distribution of locked oscillators over the branches
remains close to a constant value in the center of the $\omega$ range, however, 
it drops significantly close to the boundaries of the coherent region.

The averaged lifetime $T$ of the incoherent state drastically 
increases with decrease of coupling constant $\varepsilon$, what 
is shown in the inset of Fig.~\ref{fig:fig2}(a). 
Thus, below $\varepsilon\approx 7$ it is impossible to collect 
any reasonable statistics with a finite simulation time $T_{max}=10^6$. 
However, even for relatively small values of $\varepsilon$, the transition 
from the incoherent state is possible, what is shown by the blue markers to 
the left of they gray colored area.

It is instructive to characterize this finite-size induced transition to synchrony
in dependence on the system size $N$. 
The dependence of averaged lifetime $T$ on the rescaled 
coupling $\varepsilon/N$, plotted in the Fig.~\ref{fig:fig2}(c),
demonstrates a nice collapse
of data points. This scaling follows from the fact that the characteristic amplitude
of the coupling term is $\varepsilon R^2$ and $R\sim N^{-1/2}$ in the disordered state.
Furthermore, in the inset of 
the Fig.~\ref{fig:fig2}(c) we show a rescaled order parameter, obtained at the
end of a fixed integration time $T_{max}$; one can see that it scales
as $\mean{R}\sim N^{-1/2}f(\varepsilon N)$, what can be explained as follows. 
For sufficiently small values of the coupling the system 
always exhibits finite-size fluctuations $R\sim N^{-1/2}$ 
and remains in the asynchronous state.
With increase of $\varepsilon$, the transition to synchronous 
state becomes more probable what 
leads to an increase of the averaged final value of $R$. The upper branch
reflects the scaling of the lowest border of synchronous states $R\sim \varepsilon^{-2}$
mentioned above. Note that the critical value of coupling resulting from this scaling
is $\varepsilon\sim N$, so the transition effectively disappears in the thermodynamic limit. 

Finally, we describe the finite-size-induced 
transitions to synchrony in the 
ensemble of identical oscillators ($\omega_k=0$) with noise $D\neq 0$ (\ref{eq:main}).
Without lose of generality, we can take $D=1$, so that 
the only parameters are $\varepsilon$ and $N$.
In the thermodynamic limit, when $N\to\infty$, the system does not have any 
non-trivial coherent solution, because due to the symmetry $\varphi\to\varphi+\pi$ 
of the coupling function, the stationary density $\rho$ that follows 
from the Fokker-Planck equation
is also symmetric (in the small noise limit it tends to \eqref{eq:indic} with $S=1/2$),
thus the only stationary solution is the incoherent one with $R=0$. 
However, similar to the situations described above, 
for small system sizes $N$ a transition  to  synchronous 
two-cluster configurations is observed (cf. Ref.~\cite{Pikovsky-Rateitschak-Kurths-94}).
In contradistinction to the noise-free case, here also  back transitions
to disorder are possible due to noise, 
so that at a long run the process looks like
an intermittent order-disorder dynamics.  

\begin{figure}
\centering
\includegraphics[width=0.45\columnwidth]{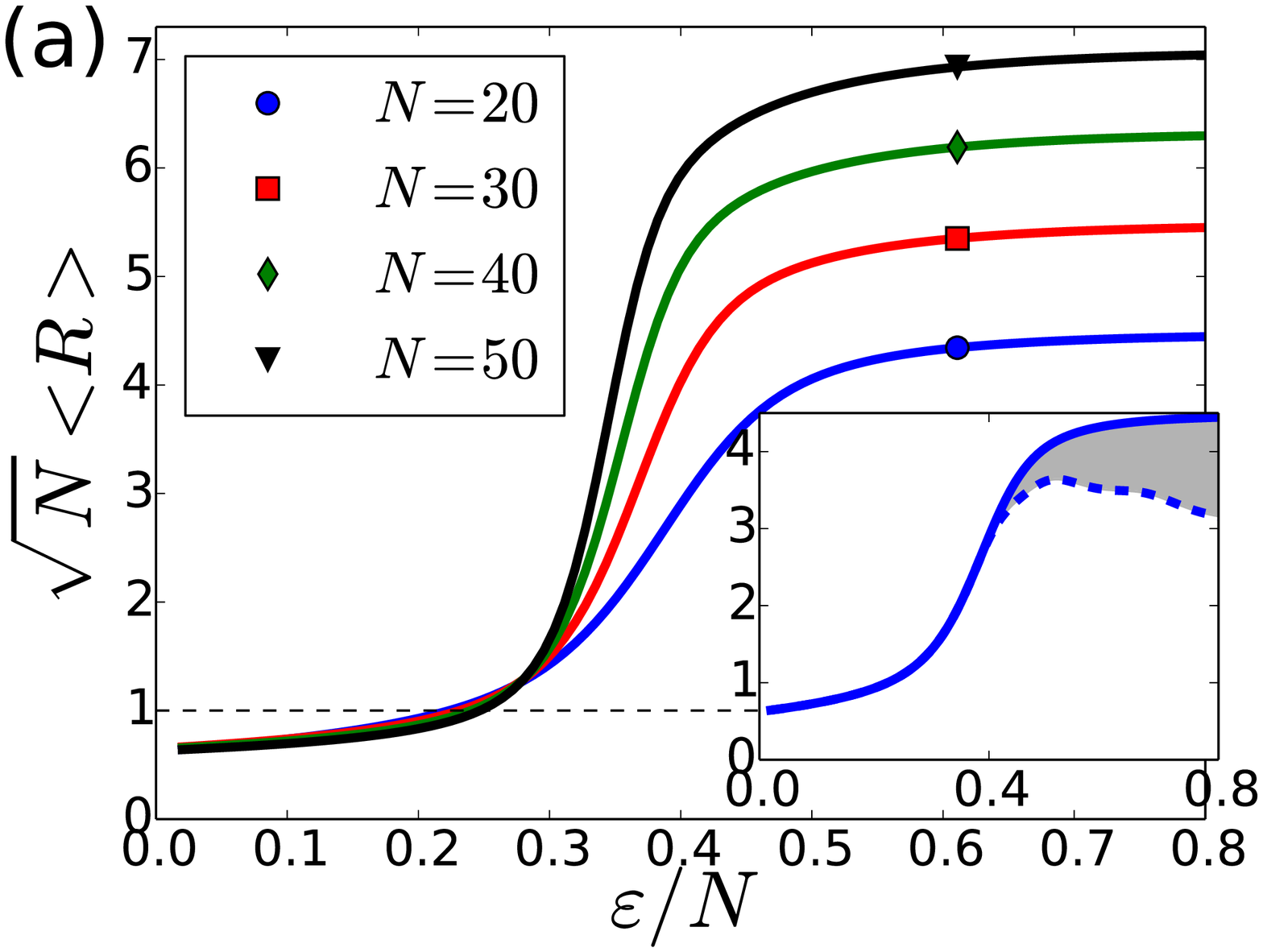}\hfill
\includegraphics[width=0.45\columnwidth]{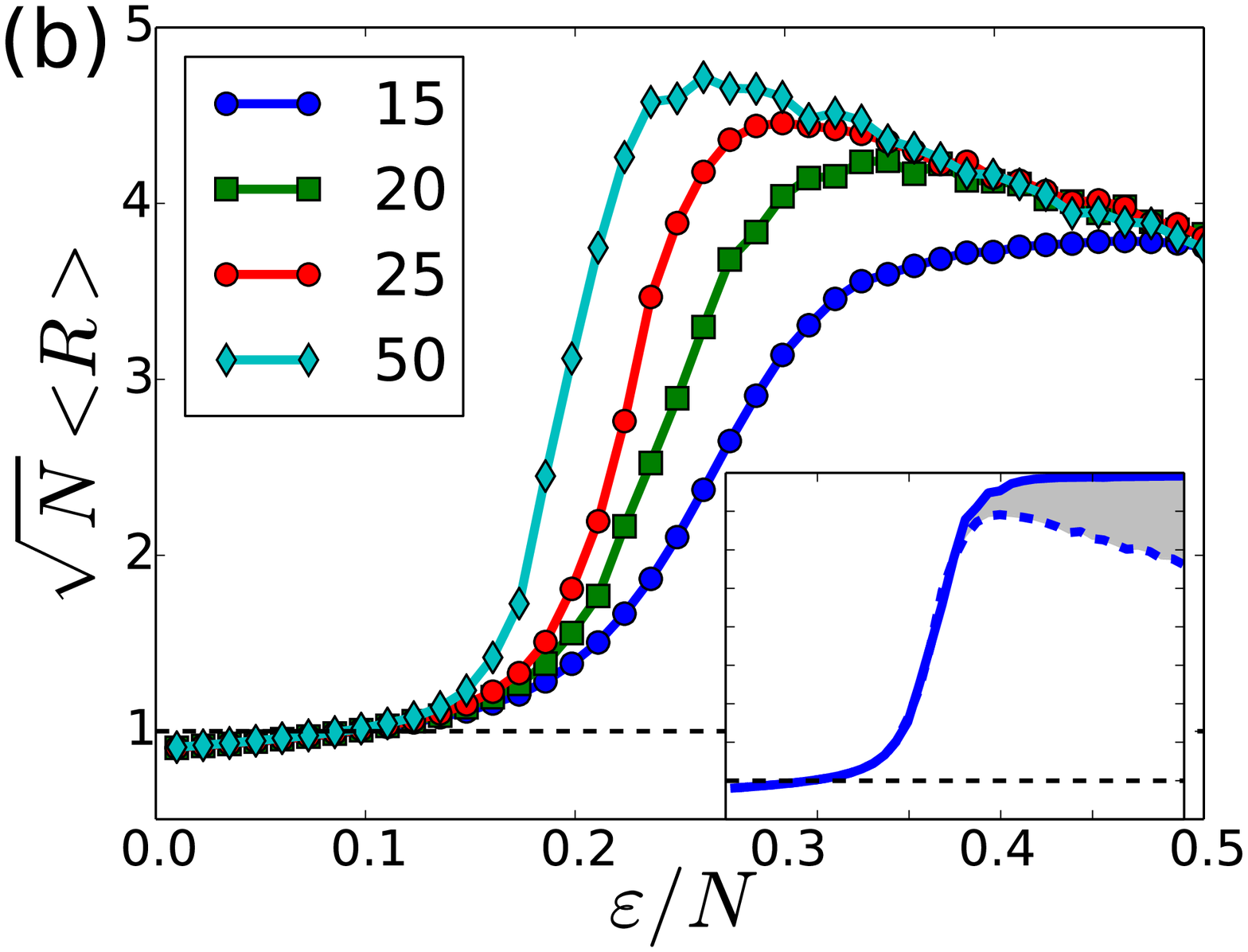}
\caption{(color online) Dependence of the order parameter $R$ on the rescaled coupling 
constant for identical oscillators with 
noise.  (a) Stationary state in the master equation as described in text.  
The inset shows the difference between
the steady state (solid line) and the one evolving from the disordered 
state at finite time (dashed line) for $N=20$. 
(b) Direct numerical simulations of ensemble ~\eqref{eq:main}
(observation time $T_{max}=10^6$). 
The inset shows results for system size 
$N=25$, but for different initial conditions (solid line: starting 
from a state with large $R$, dashed: starting from the 
disordered state).
}
\label{fig:fig3}
\end{figure}

Qualitatively, this dynamics
can be understood as effect of noise on the multiplicity of synchronous states
described for the noise-free case above (cf. discussion 
around Eq.~\eqref{eq:indic}): 
due to small noise  now transitions between these states 
$(n_1,n_2)\to(n_1\pm 1,n_2\mp 1)$
occur. The transitions rates one can estimate using the 
Kramers' formula, they are
exponentially small in the potential barrier which is 
$2\varepsilon R^2$ where $R=|2n_1/N-1|$.
For small $R$, the Kramers' rate does not apply, here one can 
phenomenologically set the transition rate
to a constant. As a result, one obtains for the order 
parameter $R$ a random walk model, which can be described 
by the corresponding master equation. Without going into details, 
which will be presented elsewhere,
we present here the main results of this statistical model. 
The stationary distribution (Fig.~\ref{fig:fig3}(a)) 
shows a transition to synchrony at $\varepsilon_c\approx 0.35 N$, 
so that for larger couplings we get $R\approx 1$,
while below this threshold only finite-size fluctuations 
around disordered state with $R\sim N^{-1/2}$ are observed.
The characteristic time scale of the time evolution from 
asynchrony to synchrony is however extremely large,
because the Kramers' rate at large $R$ is exponentially 
small. Thus, direct simulations of finite ensembles, started 
from a disordered state and
performed over a finite time interval $T_{max}$, allow to 
reveal only order parameters for which 
$\exp{\varepsilon R^2}\lesssim T_{max}$, thus $R_{max}\sim (\ln T_{max})^{1/2}\varepsilon^{-1/2}$. 
At this stage the evolution becomes 
effectively ``frozen''. We illustrate this in Fig.~\ref{fig:fig3}(b): 
only for $N\lesssim 15$ one observes a saturation
of the order parameter as predicted by the random walk model, 
while for larger $N$, values of order
parameter close to one are never achieved
during available integration times. Of course, if one starts 
from a state with $R$ close to one, it remains practically
frozen as well.

Summarizing, in this letter we considered a model of oscillators, globally coupled 
via a nonlinear function (in our case, square) of the Kuramoto mean field. Equivalently,
on the microscopic level such a coupling can be described as a 
fully connected hypernetwork. While the
disordered state remains stable in the linear approximation and 
in the thermodynamic
limit, a transition to synchrony is observed due to finite-size 
effects: the characteristic critical coupling parameter value 
scales typically as $\varepsilon\sim N$, also the transient time 
from disorder to order diverges as $N\to\infty$. 
For the deterministic 
ensembles we demonstrated  scaling properties of the transition 
in form of dependence of the order parameter 
on the coupling strength and the ensemble size. For the noisy case, 
the system demonstrates effective breaking
of ergodicity, being trapped in frozen metastable states due to 
exponentially small hopping rates. While we focused
on the purely quadratic in mean field coupling, the described 
framework allows one also to consider a general
combination of linear and nonlinear couplings.
The approach based on the master equation 
provides a framework for a description of 
 finite-size transitions not only in 
the context of phase oscillator networks, but in other types of 
mean-field coupled
systems demonstrating finite-size-induced transition~\cite{Pikovsky-Rateitschak-Kurths-94}.

\acknowledgments
M.K. thanks Alexander von Humboldt foundation,  NIH (R01 DC012943), ONR (N000141310672) and the Russian Science Foundation (Project 14-12-00811)
 for support. 
A. P. was supported by COSMOS ITN (EU Horizon 2020 research and innovation programme
under Maria-Sklodowska-Curie grant agreement No 642563) and by 
the grant (agreement 02.В.49.21.0003 of August 27, 2013  between the 
Russian Ministry of Education and Science and Lobachevsky State University of
Nizhni Novgorod).

\end{document}